\RequirePackage{fix-cm}
\documentclass[smallextended]{svjour3}       
\smartqed  
\usepackage{graphicx}
 \usepackage{subfigure}    

\usepackage{amsmath}
\usepackage{amssymb}
\usepackage{epsfig,multirow}
\usepackage{mathrsfs}
\usepackage{amsfonts}

\usepackage{xcolor}

\newtheorem{algo}{Algorithm}

\newcommand{\beq}{\begin{equation}}
\newcommand{\eeq}{\end{equation}}
\newcommand{\bay}{\begin{array}}
\newcommand{\eay}{\end{array}}
\newcommand{\bea}{\begin{eqnarray}}
\newcommand{\eea}{\end{eqnarray}}
\newcommand{\beaa}{\begin{eqnarray*}}
\newcommand{\eeaa}{\end{eqnarray*}}

\def\bdes{\begin{description}}
\def\edes{\end{description}}
\def\benu{\begin{enumerate}}
\def\eenu{\end{enumerate}}
\def\bitm{\begin{itemize}}
\def\eitm{\end{itemize}}

\newcommand{\reff}[1]{(\ref{#1})}

\newcommand{\reft}[1]{Theorem~\ref{#1}}

\newcommand{\refp}[1]{Proposition\ \ref{#1}}

\newcommand{\refal}[1]{Algorithm\ \ref{#1}}

\def\al{\alpha}

\def\argmin{\mathop{{\rm argmin}\,}}

\def\b{\bf}

\def\be{\beta}
\def\beaa{\begin{eqnarray*}}
\def\beq{\begin{equation}}
\def\beql#1{\begin{equation}\label{#1}}
\def\ba{\bar}
\def\bsl{\begin{slide}}

\def\cC{{\cal C}}\def\cG{{\cal G}}\def\cN{{\cal N}}\def\cP{{\cal P}}
\def\cQ{{\cal Q}}\def\cR{{\cal R}}\def\cS{{\cal S}}\def\cU{{\cal U}}\def\cV{{\cal V}}\def\cL{{\cal L}}

\def\cd{\cdot}

\def\cPP{{\t\cP}}

\def\cvar{{\rm CVaR\,}}

\def\E{\mathbb{E}}

\def\esssup{{\rm esup\,}}\def\essinf{{\rm einf\,}}
\def\E{\mathbb{E}}
\def\endproof{\hfill \quad{$\Box$}\smallskip}

\def\eeaa{\end{eqnarray*}}
\def\eeq{\end{equation}}

\def\esl{\end{slide}}
\def\et{\eta}

\def\cF{{\cal F}}
\def\fo{\forall}

\def\h{\hbox}
\def\ha{\hat}
\def\i{\infty}

\def\LL{\sL}
\def\l{\left}

\def\lan{\left\langle}

\def\LR{\ \Longleftrightarrow\ }
\def\MM{{\cal M}}

\def\n{\noindent}
\def\NN{{\cal N}}

\def\QQ{\mathcal{Q}}\def\QQQQ{{\t\cQ}}

\def\ov{\over}

\def\pa{\partial}

\def\proof{\noindent{\bf Proof. }}

\def\Q{{\eta}}

\def\ran{\right\rangle}

\def\r{\right}

\def\ri{{\rm ri}\,}
\def\R{\mathbb{R}}

\def\sH{\mathscr{H}}

\def\sL{\mathscr{L}}

\def\su{\subset}

\def\t{\tilde}
\def\ti{\times}

\def\v{\vskip 12 pt}

\def\VV{{\cal V}}

\def\X{{\ze}}

\def\ze{\zeta}

\journalname{Mathematical Programming, Series B}

\begin{document}

\title{Risk Minimization, Regret Minimization and Progressive Hedging Algorithms}

\titlerunning{Risk Minimization, Regret Minimization and PHAs}        

\author{Jie Sun \and Xinmin Yang \and Qiang Yao \and Min Zhang}


\institute{J. Sun \at
              Faculty of Science and Engineering, Curtin University, Australia, and \\
              School of Business, National University of Singapore\\
              \email{jie.sun@curtin.edu.au}
           \and
           X. Yang \at
           School of Mathematical Science, Chongqing Normal University, PRC\\
           \email{xmyang@cqnu.edu.cn}
           \and
           Q. Yao \at
           School of Statistics, East China Normal University, PRC, and \\
           NYU-ECNU Institute of Mathematical Sciences at NYU Shanghai, PRC\\
           \email{qyao@sfs.ecnu.edu.cn}
           \and
           M. Zhang \at
           State Key Laboratory of Desert and Oasis Ecology, Xinjiang Institute of Ecology and Geography, Chinese Academy of Sciences, PRC, and\\
           University of Chinese Academy of Science, PRC\\
           \email{zhangmin1206@ms.xjb.ac.cn}
}

\date{Received: date / Accepted: date}

\maketitle

\begin{abstract}
 This paper begins with a study on the dual representations of risk and regret measures and their impact on modeling multistage decision making under uncertainty.  A relationship between risk envelopes and regret envelopes is established by using the Lagrangian duality theory. Such a relationship opens a door  to a decomposition scheme, called progressive hedging, for solving multistage risk minimization and regret minimization problems. In particular, the classical progressive hedging algorithm is modified in order to handle a new class of linkage constraints that arises from  reformulations and other applications of risk and regret minimization models.  Numerical results are provided to show the efficiency of the progressive hedging  algorithms.
\keywords{Progressive hedging algorithm \and regret minimization \and risk measures \and stochastic optimization}
\subclass{90C15\and 90C25\and 90C34}
\end{abstract}


\section{Introduction}\label{intro}

At the core of stochastic optimization is the problem of minimizing $\E(f(x(\xi),\xi)),$ where  $\xi $ is an $m$-dimensional random vector.
For each realization (i.e., scenario) of $\xi$, $f(\cd,\xi):\R^n\to(-\i,+\i]$ is a multistage closed cost function, $x(\xi)\in\R^n$ is a decision vector (which is naturally dependent on $\xi$),    and $\E$ stands for the expectation. For the applications and algorithms we are concerned, we may assume that $\xi\in\sL^2_m$, the Hilbert space of  $m$-dimensional random vectors with support $\Xi.$ Correspondingly, we assume that the solution to the problem is a response function $x(\cd):\Xi\to \R^n$ and
$\cF(x(\cd)):=\E(f(x(\xi),\xi))$
is a proper closed convex functional of $x(\cd)$.

 As explained in detail in Rockafellar (2007) and Shapiro et al (2009), it is reasonable to replace $\E(f(x(\xi),\xi))$ by a more general {\em risk measure} $\cR(f(x(\xi),\xi))$, where $\cR$ is a functional defined on the 1-dimensional probability space $\sL^2_1$ ($\sL^2$ for short), in which norm and inner product are respectively defined as
$ \|\X\|_2:=\left[\E(\X^2)\right]^{{1/2}}$ and $\lan\ze,\et\ran:=\E(\ze\et).$
 For notational convenience, we shall specifically use $\et$ and $\ze$ to represent 1-dimensional random vectors, i.e., random variables, while use other Greek letters to denote general (probably, higher-dimensional) random vectors. Instead of minimizing $\E(f(x(\xi),\xi))$, we turn to the problem  \beq\label{P1}\min_{x(\cd)\in X}\cR(f(x(\xi),\xi)),\eeq  where $X$ is a certain feasible set of  $x(\cd)$. For convenience of discussion, we henceforth call  \reff{P1} the {\em risk minimization} problem. This idea can be extended to include ``risk  constraints" $\cR_i(f_i(x(\xi),\xi))\le0,i=1,...,m,$ in the definition of the feasible set $X$.

For both theoretical and practical purposes, we prefer $\cR$ to be ``coherent and averse".  A risk measure $\cR:~\LL^2 \rightarrow(-\infty,+\infty]$ is {\em coherent} if it satisfies the following axioms (Artzner et al. 1999, Rockafellar 2007).
\begin{description}
  \item[(A1)] $\cR(C)=C$ for all constant $C$ (``constancy equivalence"),
  \item[(A2)] $\cR((1-\lambda)\X+\lambda \X')\le (1-\lambda)\cR (\X)+\lambda\cR(\X')$ for $\lambda\in[0,1]$ (``convexity''),
  \item[(A3)] $\cR(\X)\le\cR(\X')$ if $\X\le\X'$ almost everywhere (``monotonicity''),
\item[(A4)]$\cR(\X)\le C$ when $\|\X^k-\X\|_2\to 0$ with $\cR(\X^k)\le C$ (``closedness''), and
  \item[(A5)] $\cR(\lambda \X)=\lambda\cR(\X) $ for $\lambda>0$ (``positive homogeneity'').
\end{description}
We say that the risk measure $\cR$ is {\em averse}, if it satisfies axioms (A1), (A2), (A4), (A5), and
\begin{description}
  \item[(A6)] $\cR(\X)>\E(\X)$ for all non-constant $\X$.
\end{description}
Following Rockafellar and Royset (2015), we say a risk measure is {\em regular } if it satisfies (A1), (A2), (A4) and (A6).

Risk minimization has an intrinsic connection with what we call {\em regret minimization}. Paired with the notion of risk measure, there is a notion of regret measure, denoted by $\cV.$  In operations research the notion of regret is associated to the notion of utility, namely, the regret is  regarded as the negative utility. That is
$$\cV(\X)=-\cU(-\X),$$
where $\cU$ is a certain utility functional of $-\X$, noting that $-\X$ is the ``gain" if $\X$ stands for the ``loss" (which is adopted throughout this paper). Therefore, all our subsequent results could have corresponding interpretations in utility models.

Similarly to coherent risk measures, we define {\em coherent regret measures} $\VV$ as follows\footnote{The notion of coherent regret measure appears to be new in the literature.}. A functional
$\VV:~\LL^2 \rightarrow(-\infty,+\infty]$ is called a coherent regret measure if it satisfies the following axioms.
\begin{description}
  \item[(B1)] $\VV(0)=0$ (``zero equivalence"),
  \item[(B2)] $\VV((1-\lambda)\X+\lambda \X')\le (1-\lambda)\VV (\X)+\lambda\VV(\X')$ for $\lambda\in[0,1]$ (``convexity''),
  \item[(B3)] $\VV(\X)\le\VV(\X')$ if $\X\le \X'$ almost everywhere (``monotonicity''),
  \item[(B4)] $\VV(\X)\le0$ when $\|\X^k-\X\|_2\to 0$ with $\VV(\X^k)\le0$ (``closedness''), and
  \item[(B5)] $\VV(\lambda \X)=\lambda\VV(\X) $ for $\lambda>0$ (``positive homogeneity'').
\end{description}
We say that the regret measure $\VV$ is {\em averse}, if it satisfies axioms (B1), (B2), (B4), (B5) and
\begin{description}
  \item[(B6)] $\VV(\X)>\E(\X)$ for all nonzero $\X$.
\end{description}
In addition, we say that a regret measure is {\em regular} if it satisfies (B1), (B2), (B4) and (B6).

In the theory of risk quadrangle of Rockafellar and Uryasev (2013), a risk measure could be understood as the ``certainty-uncertainty trade-off" of a regret measure, namely, a risk measure could be defined through a regret measure as
\begin{equation}\label{e:vtor}\cR(\X)=\inf_{y\in \R}\{y+\cV(\X-y)\},\end{equation} where $y$ is a single real variable. This formula generalizes the formula for conditional value-at-risk (CVaR for short), popularized by Rockafellar and Uryasev (2000).

Rockafellar and Royset (2015, Theorem 2.2) showed that for regular risk and regret measures, the ``inf" in \reff{e:vtor} can be replaced by ``min" because the infimum is attainable. This then implies that the same is true for coherent averse risk and regret measures. We moreover in this paper deduce a result that describes the relationship between the dual representations of risk and regret measures. Based on that result and earlier work on duality of risk measures, we determine the dual envelopes (defined later) of several popular regret measures and provide a general approach to finding the  $\cV$ in \reff{e:vtor} given $\cR$ in Section 2.

In terms of optimization, formula \reff{e:vtor} opens a door for converting a risk minimization problem to a regret minimization problem; i.e.,
\beq\label{P3}\min_{x(\cd)\in X}\cR(f(x(\xi),\xi)) \Longleftrightarrow
\min_{x(\cd)\in X, y\in \R}[y+\cV(f(x(\xi),\xi)-y)],
\eeq
which would allow to minimize a coherent averse risk measure by a decomposition approach if  $\cV$ can be expressed as an expectation of a  function of $\xi$.
Starting from Section 3, the second part of this paper  is concerned with a computational method for risk minimization \reff{P1} and the corresponding regret minimization \reff{P3}. We introduce the progressive hedging algorithm, originally developed by Rockafellar and Wets (1991) for solving multistage convex stochastic optimization problems and later extended  to solving monotone stochastic variational inequality problems (Rockafellar and Sun, 2019). We explain how this algorithm could be used to solve \reff{P1} and \reff{P3}. Since in certain circumstances the original progressive hedging algorithm cannot be used due to a new ``linkage" constraint that links different $\xi$ into one constraint, a modified progressive hedging algorithm is proposed. Compared with other algorithms for multistage stochastic optimization, say for example, the distributionally robust approach of Wiesemann et al. (2014), progressive hedging is less restrictive in the sense that  it requires no linear decision rule, can handle more general nonlinear objective functions, and is easily expandable to more than two stages. Numerical results will be presented in Section 4, where we show that the algorithm is fairly efficient for solving middle-sized  coherent and averse risk/regret minimization problems.

The first paper  working on the format of problem \reff{P3} in the context of progressive hedging algorithms is  Rockafellar (2018) that was concentrated on the CVaR measure and only briefly mentioned the general case in its last section. This paper could be thought of as a further development of Rockafellar's work  with the following contributions.

\bitm
\item We develop a dual theory for the relationship between coherent risk measures and regret measures and use it as a stepping stone in developing new models of minimizing regret measures, which may expand the applicability of stochastic optimization approaches.
\item  As certain linkage constraints may arise in the above models, these  models may not be in the required decomposable format of the progressive hedging algorithm. We modify the progressive hedging algorithm for handling these cases. The modified progressive hedging algorithm takes advantage of the hidden decomposability of the problem and keeps the computational effort at the same level.
\item We provide numerical evidence to show the power of the  progressive hedging algorithms. The tested samples include problems from real applications and randomly generated ones that are more of  ``practical size", compared to the examples in the literature of progressive hedging algorithms.
\eitm

\section{ The dual representation of risk and regret measures}

It is well known (Rockafellar, 2007) that any coherent risk measure $\cR$ has a dual representation; that is, there is a nonempty, convex and weakly closed\footnote{Since for convex sets in $\sL^2$ strong closedness and weak closedness coincide, we shall not differentiate strong and weak closedness in statements on convex sets below.} set $\cQ\subset \LL^2 $,  which can be shown to be unique, called ``the risk envelope'' of $\cR$, such that for any $\X\in\LL^2$,
\begin{equation}\label{e:1r}
\cR(\X)=\sup\limits_{\Q\in\cQ}\lan\X,\Q\ran.
\end{equation}
Moreover, $\cQ$ is a subset of $$\cP:=\{\Q\in\sL^2:~\E(\Q)=1, \Q\ge0\}.$$ More detailed analysis can be seen in Ang et al. (2018) and Rockafellar and Uryasev (2013).

The dual representation for $\VV$ can be similarly established. By convex analysis (Clarke (2013), Theorem 4.25, the finite-dimensional version of it appeared in Rockafellar (1970), Theorem 13.2), any functional that satisfies (B1)-(B5) can be represented as a specific support function. That is, there is a unique, nonempty, convex, and closed $\QQQQ\su \sL^2$, such that
\begin{equation}\label{e:v1}
\VV(\X)=\sup\limits_{\Q\in\QQQQ}\lan\X,\Q\ran,
\end{equation}
where $\QQQQ$ is a subset of
$$\cPP:=\{\Q\in\sL^2:\Q\ge0\}.$$

In the next, starting with the basic equation (\ref{e:vtor}), we investigate the relationship between $\cQ$ and $\QQQQ,$ and later present several explicit descriptions of $\cQ$ and $\tilde\cQ$  for certain popular risk and regret measures.

\subsection{Relationship between risk and regret envelopes }

The next proposition formalizes the statement \reff{e:vtor} on  risk and regret measures.
\begin{proposition}\label{p:vtor}
For any risk measure $\cR$ satisfying (A1) and (A2), which includes coherent, averse, or regular risk measures as special cases, there exists at least one  regret measure $\VV$ of the same type (e.g., coherent, averse, or regular), such that (\ref{e:vtor}) is valid with $\inf$ being replaced by $\min$.
\end{proposition}
\proof Just note that Axioms (A1) and (A2) imply $y+\cR(\ze-y)=\cR(y+\ze-y)$ (Rockafellar et al, 2006) and $\cR$ itself can be a candidate for $\VV$ to satisfy (\ref{e:vtor}). \endproof

\noindent\textbf{Remarks.}
 \begin{itemize}
\item[(i)]  It should be noted that the opposite of \refp{p:vtor} is not true; for instance, even if $\VV$ is a  coherent regret measure, the functional $\cR$ obtained via (\ref{e:vtor}) may not be a risk measure. Say, if $\VV(\X)=2\E(\X)$, then $\VV$ is a  coherent regret measure by direct verification of (B1)-(B5), but the functional $\cR$ obtained via (\ref{e:vtor}) $\equiv-\infty$, which indicates that this $\cR$ is not a risk measure. Hence it is important to find the conditions for $\VV$ to guarantee $\cR$ defined via (\ref{e:vtor}) to be an appropriate risk measure.

\item[(ii)] For a given coherent risk measure $\cR$, there may be more than one  $\VV$ satisfying relationship (\ref{e:vtor}). For instance, let $\VV_1(\X)=\E(\X_+)$ and $\VV_2(\X)=\E(\X_+)+\E(\X),$ where $\X_+:=\max (\X,0)$. Then it can be shown that $\cR_1(\X)=\cR_2(\X)=\E(\X)$ (see Section 2.2.2).
 \end{itemize}

\n The next theorem establishes the relationship among coherent $\cR$,  $\VV$, $\QQ$, and $\QQQQ$.
\begin{theorem}\label{t:main}
Suppose that $\cR$ is a coherent risk measure with the dual representation (\ref{e:1r}) and $\VV$ is a  coherent regret measure with the dual representation (\ref{e:v1}), where $\QQQQ$ is weakly compact.
 Then $\cR$ and $\VV$ satisfy relationship (\ref{e:vtor})   if and only if $\QQ=\QQQQ\cap\cP$.
\end{theorem}
\proof Fix $\X\in\LL^2$, let $$L(\Q,y):=\E(\X\Q)+y[1-\E(\Q)]
$$ for $y\in\R$ and $\Q\in\QQQQ$. From weak compactness of $\QQQQ$ and the Fan minimax theorem (Fan, 1953, Theorem 2), we  have
\begin{equation}\label{e:maxmineq}
\sup_{\Q\in\QQQQ}\inf_{y\in\R}L(\Q,y)=\inf_{y\in\R}\sup_{\Q\in\QQQQ}L(\Q,y).
\end{equation}
Since $$\inf_{y\in\R}L(\Q,y)=\left\{\begin{array}{ll}\E(\X\Q),~~~~~~\text{if}~\E(\et)=1,\\
~-\infty,~~~~~~~~\text{otherwise}, \end{array}\right.$$
we have
\begin{equation}\label{61}
\sup_{\Q\in\QQQQ}\inf_{y\in\R}L(\Q,y)=\sup_{\Q\in\QQQQ\cap\cP}\E(\X\Q).
\end{equation}
Notice that by (\ref{e:v1}) and the definition of $L(\Q,y)$, we have
\begin{equation}\label{71}
\inf_{y\in\R}\sup_{\Q\in\QQQQ}L(\Q,y)=\inf_{y\in\R}\{y+\VV(\X-y)\},\hbox{ where }\VV(\X)=\sup_{\Q\in\QQQQ}\E(\X\Q).
\end{equation}
Thus by (\ref{e:maxmineq}),\reff{61}, and \reff{71}, we obtain
\begin{equation}\label{e:maxmineq1}
\sup_{\Q\in\QQQQ\cap\cP}\E(\X\Q)=\inf_{y\in\R}\{y+\VV(\X-y)\}.
\end{equation}

If $\QQ=\QQQQ\cap\cP$, then by (\ref{e:maxmineq1}) we have
$$\sup\limits_{\Q\in\QQ}\E(\X\Q)=\inf\limits_{y\in\R}\{y+\VV(\X-y)\}.$$ In view of (\ref{e:1r}), it follows that the relationship (\ref{e:vtor}) holds for $\cR$ and $\VV$.

Conversely, if $\cR$ and $\VV$ have relationship (\ref{e:vtor}), then by (\ref{e:vtor}) and (\ref{e:maxmineq1}) we get
$$\cR(\X)=\sup\limits_{\Q\in\QQQQ\cap\cP}\E(\X\Q).$$  It is easy to see that $\QQQQ\cap\cP$ is a nonempty, convex and closed subset of $\LL^2$, and therefore, it is a risk envelope of $\cR$. By the uniqueness of risk envelope, we have $\QQ=\QQQQ\cap\cP$.\endproof

Theorem \ref{t:main} has some overlapping with Theorem 2.2 in Rockafellar and Royset (2015), as well as the Envelope Theorem in Rockafellar and Uryasev (2013). However, the relationship on the two envelopes appears to be new. Note that the result is derived by an elementary   approach without involving the conjugate function theory in paired space (Rockafellar, 1966, 1974). Theorem \ref{t:main} provides a way to determine a coherent regret measure $\VV$ corresponding to a given coherent risk measure $\cR$ as follows. Given a coherent risk measure $\cR$, find its risk envelope $\QQ$, and relax the condition ``$\E(\Q)=1$'' to get $\QQQQ$, then (\ref{e:v1}) determines the corresponding $\VV$. Note that since there may be more than one way to relax the condition ``$\E(\Q)=1$'', there may be more than one $\VV$ corresponding to the same $\cR$ as well.  

A complication for applying Theorem 1 is the requirement of weak compactness for $\t\cQ$. Since under the one-to-one correspondence between nonempty closed convex sets in $\sL^2$ and closed positive homogeneous
convex functions on $\sL^2$ (Rockafellar, 1966), the set $\t\cQ$ is weakly compact if and only if the function $\cV$ is continuous everywhere.   Thus,
this obstacle can be removed by checking the global continuity of $\cV$ on $\sL^2,$ which is specifically true if the random variable has a finite discrete distribution (Rockafellar and Uryasev, 2013) as in the case when the progressive hedging algorithm is considered. We write this observation as a corollary below.
\begin{corollary}\label{s1}
\reft{t:main} remains to be true if the condition ``$\t\cQ$ is weakly compact" is replaced by ``$\cV$ is continuous everywhere".
\end{corollary}

\subsection{Examples of popular $\QQ$ and $\QQQQ$}

This subsection provides examples of popular pairs of risk and regret measures. Some of the conclusions have appeared in Rockafellar and Uryasev (2013) and Ang et.al. (2018), but no detail of the  regret envelopes was given before. We also display the respective risk  and regret functions for the purpose of algorithmic development in the following sections.

\subsubsection{Optimized certainty equivalence (OCE) and CVaR}\label{sub:OCE}

Given $0\leq\gamma_2<1\leq\gamma_1$, let $\cal O$ be the OCE-measure introduced by Ben-Tal and Teboulle (2007).
$${\cal O}(\X):=\inf_{y \in \R}\{ y+\E(r(\ze- y))\},$$
where
$r(\X)=\gamma_1 \X_+-\gamma_2 \X_-$ with $\X_-:=\min(\X,0)$ and $\X_+:=\max(\X,0).$

It is shown in Ang et al. (2018) that the OCE-measure is a coherent risk measure with risk envelope $$\QQ_{\gamma_1,\gamma_2}=\left\{\Q:~\gamma_2\leq \Q\leq\gamma_1,~\E(\Q)=1\right\}.$$
Removing the condition ``$\E(\Q)=1$'', we get a  set $$\QQQQ_{\gamma_1,\gamma_2}=\left\{\Q:~0\le \gamma_2\leq \Q\leq\gamma_1\right\}.$$ Therefore, the corresponding regret measure is $$\VV_{\gamma_1,\gamma_2}(\X)=\sup\limits_{\Q\in\QQQQ_{\gamma_1,\gamma_2}}\E(\X\Q)=\gamma_1\E({\X}_+)-\gamma_2\E({\X}_-).$$ Note that $\VV_{\gamma_1,\gamma_2}$ is finite and convex on $\sL^2$, so it is continuous everywhere. Therefore $\t\cQ$ is weakly compact and Theorem 1 is applicable here.

In particular, if we take $\gamma_1=(1-\alpha)^{-1}$ and $\gamma_2=0$, where $0\leq\alpha<1$, then the OCE-measure becomes the  measure of $\cvar_\alpha$. The corresponding regret measure  is $$\VV_\alpha(\X)=\frac{1}{1-\alpha}\E({\X}_+)\h{ with } {\t Q}_\al=\left\{\Q:~0\leq \Q\leq{1\ov1-\al}\right\}.$$
Then formula \reff{e:vtor} is in fact the ``minimization formula'' of CVaR, i.e., $$\cvar_\alpha(\X)=\min\limits_{y\in\R}\left\{y+\frac{1}{1-\alpha}\E(\X-y)_+\right\},$$ which is a consequence of \reft{t:main}.

It is interesting to observe that OCE-measure is representable by CVaR, namely
$\cal O(\X)=\gamma_2\E(\X)+\cvar_\alpha(\X),$ where $\alpha=1-(\gamma_1-\gamma_2)^{-1}.$ Thus OCE-measure and CVaR are in a sense equivalent.

\subsubsection{Expectation as risk measure}\label{sub:expectation}

This is a special case of CVaR when $\alpha=0$ and $\QQ=\{1\}$. That is, $$\cR(\X)=\E(\X).$$
By the result of Section \ref{sub:OCE}, a candidate for the corresponding regret measure is $$\VV(\X)=\E({\X}_+).$$
On the other hand, since $\overline{\QQ}=\{\Q:~1\leq \Q\leq 2\}$ also satisfies $\overline{\QQ}\cap\cP=\{1\}$, and it is bounded in $\LL^2$, we find that
\beaa
\overline{\VV}(\X)&=&\sup\limits_{1\leq \Q\leq 2}\E(\X\Q)\\&=&\sup\limits_{0\leq \Q\leq 1}\E(\X+\X\Q)\\
&=&\E(\X)+\E({\X}_+)
\eeaa is another candidate for the corresponding regret measure. Since both $\overline\cV$ and $\cV$ are continuous on $\sL^2$, both $\overline\cQ$ and $\t\cQ$ are weakly compact, it is valid to apply Theorem 1 in these cases.

\subsubsection{Worst case as risk measure}\label{sub:worstcase}

This risk measure is defined as $$\cR(\X)=\esssup(\X),$$where $\esssup$ is the essential-sup function ($\essinf$is similarly defined). Note that the worst case risk measure may not be finite and the corresponding risk envelope $\QQ=\cP$ is not bounded. However we can directly verify that \reft{t:main} is still true with $\QQQQ={\tilde \cP},$ and,
$$\VV(\X)=\left\{\begin{array}{ll}~0,~~~~~~~~~\text{if}~\X\leq0~\text{almost surely},\\+\infty,~~~~~~\text{otherwise}.\end{array}\right.$$

\subsubsection{Mean-deviation-penalty risk measure}\label{sub:meandeviation}

Fix $0\leq\lambda\leq1$. Define the Mean-deviation-penalty risk measure as$$\cR(\X)=\E (\X)+\lambda\|(\X-\E (\X))_+\|_2$$ for all $\X\in\LL^2$. From Ang et al. (2018), we know that $\cR$ is a coherent and averse risk measure with risk envelope
\begin{equation}\label{e:envelopemd}
\QQ=\left\{\Q:~\Q\geq 0,~\E(\Q)=1,~\|\Q-\essinf \Q\|_2\leq\lambda\right\}.
\end{equation}
We next find the corresponding coherent regret measure $\VV$ for it. Note that by simply getting rid of the restriction ``$\E(\Q)=1$'', we may get an unbounded subset of $\LL^2$ and therefore may get a non-finite $\VV$. To avoid it, note that $\Q\geq0$ and $\E(\Q)=1$ together imply $0\leq\essinf \Q\leq1$. Therefore, \beql{QQQQ}\QQQQ=\left\{\Q:~0\leq\essinf \Q\leq1,~\|\Q-\essinf \Q\|_2\leq\lambda\right\}\eeq is bounded and satisfies $\QQQQ\cap\cP=\QQ$. Thus, we prefer to use $\QQQQ$ for calculating $\VV(\cdot)$.

For any $\X\in\LL^2$ and $\Q\in\QQQQ$, we have
$$\E(\X\Q)\leq\E({\X}_+(\Q-\essinf \Q))+\essinf \Q\cdot\E (\X)\leq\lambda\|{\X}_+\|_2+(\E (\X))_+.$$
Furthermore, the equation holds when $\Q=\textbf{1}_{\{\E (\X)\geq0\}}+\dfrac{\lambda {\X}_+}{\|{\X}_+\|_2}$ ($0/0$ is defined as $0$). Therefore,
\beql{VV}\VV(\X)=\lambda\|{\X}_+\|_2+(\E (\X))_+\eeq is a candidate for the regret measure corresponding to the mean-deviation-penalty risk measure. The global continuity of $\VV$ guarantees the weak compactness of $\t\cQ$. Hence Theorem 1 is applicable.

We may check  \reft{t:main} for this case directly. For $y\in\R$, we have
\beaa y+\VV(\X-y)&&=y+(\E (\X)-y)_++\lambda\|(\X-y)_+\|_2\\&&=\left\{\begin{array}{ll}y+\lambda\|(\X-y)_+\|_2, & ~~~\text{if}~y\geq\E (\X),\\
\E (\X)+\lambda\|(\X-y)_+\|_2, & ~~~\text{if}~y<\E (\X).\end{array}\right.
\eeaa
Therefore, $y+\VV(\X-y)$ is decreasing in $y$ when $y<\E (\X)$ and increasing in $y$ when $y\geq\E (\X)$, and so it reaches its minimum when $y=\E (\X)$. Then (\ref{e:vtor})  holds for mean-deviation-penalty risk measure with $\QQQQ$ and $\VV$ being specified by \reff{QQQQ} and \reff{VV}, respectively.

\subsection{Aversity }\label{sub:relationship}

Aversity of risk measures is important in many applications of risk minimization. It was proven in Ang et al (2018) that a sufficient condition for aversity is that $\{{\b 1}\}$ is a relative interior point of $\QQQQ$ with respect to the plane $\{\Q:\E(\Q)=1\}$ and this condition is also necessary if the probability space is finite. As is shown in Ang et al. (2018), all of the risk measures discussed above except the expectation are coherent and averse.  By Theorem 2.2 of Rockafellar and Royset (2015), the OCE, $\cvar_\al$ ($0<\al<1$), worst case and mean-deviation-penalty regret measures are also averse.

\subsection{About max and convex combination of risk and regret measures}

In Rockafellar and Uryasev (2013), as well as in Rockafellar and Royset (2015), it was shown that some properties of risk measure  and regret measure, such as regularity, convexity, monotonicity, and positive homogeneity can be derived from each other. However, some other properties are not preserved between these two measures. Here are two examples.
In Section \ref{sub:expectation}, we proved that  regret measures $\E({\X}_+)$ and $\E (\X)+\E({\X}_+)$ correspond to the same risk measure $\E (\X)$. Note that  the maximum of the two regret measures is $$\max\{\E({\X}_+),\E(\X)+\E({\X}_+)\}=(\E (\X))_++\E({\X}_+),$$ which turns out to correspond to the generalized mean-deviation-penalty risk measure of $\lambda=1$ mentioned in  Section \ref{sub:meandeviation}.

This example demonstrates  that the maximum of several regret measures may not generate the maximum of the respective risk measures. 

We further notice that the relationship of convex combination  is not preserved, either. The following example demonstrates this point.

Let $\VV_1(\X):=\E({\X}_+)$ and $\VV_2(\X):=\E(\X)+\E({\X}_+)$ be two regret measures. Fix $0<\lambda<1$ and let $$\VV(\X):=(1-\lambda)\VV_1(\X)+\lambda\VV_2(\X)=\lambda\E(\X)+\E({\X}_+).$$ It is known that $\VV_1(\X)$ and $\VV_2(\X)$ correspond to the same risk measure $\E(\X)$. Next, we calculate the risk measure corresponding to $\VV(\X)$. It is easy to see that $$y+\VV(\X-y)=\lambda\E (\X)+(1-\lambda)y+\E((\X-y)_+).$$ Hence we have
\begin{align*}
\min\limits_{y\in\R}\{y+\VV(\X-y)\}&=\lambda\E (\X)+(1-\lambda)\min\limits_{y\in\R}\left\{y+\frac{1}{1-\lambda}\E((\X-y)_+)\right\}\\
&=\lambda\E (\X)+(1-\lambda)\cvar_\lambda(\X).
\end{align*}
Thus, the risk measure generated by $\lambda\VV_1+(1-\lambda)\VV_2$ through (\ref{e:vtor}) is not $\lambda\cR_1+(1-\lambda)\cR_2=\E(\X)$. This example shows that the convex combination relationship  does not pass from regret measures to the corresponding risk measures.

\section{The progressive hedging algorithm for risk/regret minimization}

\subsection{A multistage perspective of risk/regret  minimization}

The next focal point of this paper is the application of the progressive hedging algorithm (PHA) for the multistage risk and regret minimization. Let the objective function of \reff{P3} be the risk or regret measure of total $N$ decision periods in the form $\E_\xi(f(x(\xi),\xi))$, and let  $\Xi$ consist of a finite number of scenarios. We
aim at determining an optimal response function
 in the form of
$$
   x(\cdot)=(x_1(\xi),\ldots,x_N(\xi))^T \in
          \R^{n_1}\times\cdots\times\R^{n_N}=\R^n,
$$where ``$T$" stands for the transpose.
Let $\sH_n$ be the space of all such functions $x(\cdot)$   endowed with the expectation inner product
\beql{1.2}
   \langle x(\cdot),w(\cdot)\rangle := \E(x(\xi)\cdot w(\xi))
      = \sum_{\xi\in\Xi}p(\xi)\sum_{k=1}^N x_k(\xi)\cdot w_k(\xi),
\eeq
which makes $\sH_n$ into a finite-dimensional Hilbert space.

The multistage nature of the minimization problems requires that $\xi$ be disclosed gradually in the form $\xi=(\xi_1,\ldots,\xi_{N})$, where $\xi_i$ is revealed only after  $x_{i-1}(\xi)$ is made, but before $x_i(\xi)$ is determined.
A consequence of this fact is that the mappings $x(\cdot)$ must be {\em nonanticipative}
in the sense that  $x(\cdot)$ belongs to the so-called  nonanticipativity subspace $\cN$ of $\sH_n$, where
$$
    \cN := \l\{ x(\cdot)\in\sH_n : \begin{array}{l} \forall~k,~x_k(\xi_1,\ldots,\xi_{k-1},
       \xi_k,\ldots,\xi_N) \\
       \quad \hbox{ doesn't depend on } \xi_k,\ldots,\xi_N\end{array} \r\}.
$$
 In addition, suppose every decision $x(\cdot)$ must satisfy a separate set of constraints and these constraints generally depend on $\xi.$ We write this fact in the form of
\begin{eqnarray*}
x(\cdot)\in\mathcal{C} \subset \sH_n,~\textrm{which~means}~x(\xi)\in C(\xi) ~\forall\xi\in\Xi,
\end{eqnarray*}
where each $C(\xi)$ refers to a nonempty closed convex subset of $\mathbb{R}^{n}$ and the set $\mathcal{C}$ therefore  denotes a nonempty closed convex subset of  $\sH_n$.

We are now ready to clarify the exact meaning of the regret minimization \reff{P3} in the multistage setting. Let $X=\cN\cap\cC$ be the feasible set of \reff{P3}, consisting of all response functions $x(\cd)\in\sH_n$ that are  nonanticipative and satisfy constraint $x(\cd)\in \cC.$ Let
$$z(\cd)=(x(\cd),y)\in \ba\cL:= \sH_n\ti\R,$$and $\cG(z(\cd)):\ba\cL\to\ba\cL$ be the mapping specified by $$\cG(z(\xi))=y+\cV(f(x(\xi),\xi)-y).$$
Since $y$ is independent of $\xi$ and $x_1$ is also independent of $\xi$ under nonanticipativity, we may regard  $(x_1,y)$ as $\ba x_1$. Let $\ba\cN$ be the nonanticipativity subspace $\ba \cL$ and $\ba\cC=\R\ti\cC$. Then the regret minimization \reff{P3} becomes
\beql{SVI}
     \min_{z(\cd)}\ \cG(z(\cdot)) \h{ over all } z(\cdot)\in\ba\cC\cap{\ba\cN}.
\eeq

The progressive hedging algorithm developed by Rockafellar and Wets (1991) aims at the expectation form of $\cG(z(\cd))$, namely $\cG(z(\cd))=\E(g(z(\xi),\xi)),$
where $g(z(\xi),\xi)$ is proper, closed, and convex in $z(\xi)$ on $\ba C(\xi)$ for each $\xi\in\Xi$.

\begin{algo}\label{A31} {\b The PHA for problem \reff{SVI} with {\b $\cG(z(\cd))=\E(g(z(\xi),\xi))$}}
\v
\n{\b Step 1 (Scenario Decomposition).} Given a primal-dual  pair $z^k(\cd)\in \ba\NN$ and  $v^k(\cd)\in\ba\MM:=\ba\cN^\perp$, solving an augmented Lagrange problem for each $\xi\in\Xi$ as follows to determine ${\ha z}^k(\cd).$
\beql{2.31}
      \hat z^k(\xi) =\argmin_{z\in \bar{C}(\xi)}\!\l\{ g(z,\xi)-
   v^k(\xi)\cdot z+\frac{r}{2}||z-z^k(\xi)||^2\r\}.
\eeq Note that the vector ${\ha z}^k(\xi)$ in \reff{2.31} exists and is uniquely determined
because the proximal term forces the function being minimized to be strongly
convex.

\n{\b Step 2 (Primal and Dual Update).}
   \beql{2.32} z^{k+1}(\cdot)=P_{\ba\NN}(\hat z^k(\cdot)) \h{ and}~
    v^{k+1}(\cdot) = v^k(\cdot)-r P_{\ba\MM}(\hat z^k(\cdot)),\eeq
where $P_{\ba\NN}$ and $P_{\ba\MM}$ are the projection operators to the subspaces $\ba\NN$ and $\ba\MM$, respectively, and $r>0$ is a suitably chosen parameter. The primal update involves computation of a conditional expectation and the dual update is a simple move in the space ${\ba\MM}$. See details in Rockafellar and Wets (1991).
\end{algo}

A key advantage of PHA is the decomposability in terms of $\xi$ in Step 1. Note that in Step 1 the solution of  $\ha z^k(\xi)$ can be found in parallel on  $\xi$. The aggregated  $\hat z^k(\xi)$ becomes the $ \hat z^k(\cd)$ for Step 2. In a nutshell, finding a solution $z(\cd)$ of \reff{SVI} is generally a difficult task due to the huge dimension of $z(\cd)$. 
On the other hand, to find a solution to \reff{2.31} is much easier, which amounts to solving a strongly convex program of dimension $O(n)$. In case of $ g(z(\xi),\xi)$ being convex quadratic in $z(\xi)$, problem \reff{2.31} is a convex quadratic program and can be solved by a state-of-art package.

A few words on the convergence properties of PHA are in order. It is shown in Rockafellar and Wets (1991) that \refal{A31} generates a convergent sequence to a solution to problem \reff{SVI}, as long as \reff{SVI} is a convex problem with constraint qualification and has a solution. If in addition the sets $C(\xi)$ are polyhedra, and the mapping $\partial\cG$ is monotone and piecewise polyhedral, the rate of convergence is linear with respect to the norm
$$\|(z(\cd),v(\cd))\|_r:=\l(\|z(\cd)\|^2+r^{-2}\|v(\cd)\|^2\r)^{1/2}.$$
According to Proposition 2.2.4 of Sun (1986), in a finite dimensional Hilbert space, $\pa \cG$ is piecewise polyhedral if and only if $\cG$ is closed and convex piecewise quadratic  (including convex piecewise linear as a special case).\footnote{ A function is convex piecewise quadratic if it is convex and its domain is a union of convex polyhedra, on each of which the function is quadratic.}   In fact, \refal{A31}  converges for general convex $\cG(z(\cd))=\E(g(z(\xi),\xi))$ under assumptions on constraint qualification and existence of solution.

The choice of applying PHA to risk or regret model, i.e., to solve problem \reff{P1} or \reff{P3}, provides flexibility in practice as long as the objective function is expressible as an expectation of a convex function. However, it often happens that the regret model has a simpler form. For instance, the so-called rate-based measure (Rockafellar and Uryasev, 2013)
has
$$ \cR(\X)=r(\X)+\E \l(\log \frac{1}{1-\X+r(\X)}\r)~\h{and}~\VV(\X)=\E\l(\log\frac{1}{1-\X}\r),$$
      where $r(\X)$ is the unique $C\geq \esssup \X-1$ such that $\E((1-\X+C)^{-1})=1$.
In this case,  it   meets the requirement for objective function of PHA. Hence if other requirements for convergence are satisfied, the problem may be solvable by PHA. This example also says there are cases where the risk measure is not in the form of expectation, therefore can not be solved by PHA directly, while the corresponding regret measure is in expectation form and can be suitable for directly applying PHA.

It should be noted that \refal{A31} also requires the constraints to be of the form $z(\xi)\in \bar{C}(\xi),~\forall \xi$. This will exclude the constraint, say for example,
\beq\label{18}\E\big[h(z(\xi),\xi))\big]\le t,\eeq since this constraint involves all scenarios rather than a single scenario $\xi$. Therefore it can not be regarded as $z(\xi)\in \bar{C}(\xi)$ for certain $\bar{C}(\xi)$. Let us call this type of constraints {\it linkage constraints}. Since linkage constraints arise frequently in regret minimization, we need to remove this obstacle by certain modification of \refal{A31}, as we shall do below.

\subsection{ Occasions where a linkage constraint arises}

{\b Reformulation.} Some regret measures such as the mean-deviation-penalty may contain a single term such as $(\E(h(x(\xi),\xi)))_+$, either in the objective function or on the left-hand side of a constraint, the latter case happens particularly if the problem contains a CVaR constraint such as $\cvar_\al[(h(x(\xi),\xi)]\le\be$. In this case it may be convenient to introduce a non-random variable $t$ and a linkage constraint     \beq\label{cross}t\ge0\h{ and }t\ge \E(h(x(\xi),\xi))\eeq to replace this term in the new formulation for the purpose of simplifying the computation.
 Notice that in this case we obtain a convex linkage constraint if $h(\cd,\xi)$ is convex.

Other examples of linkage constraints can be found in early work on stochastic variational inequality, e.g., the expected residual method of Chen and Fukushima (2005) and Chen et al. (2012), which involve constraints on the expectation of some residual functions.

\n{\b Moment constraints.} Practical applications often involve constraints on the moments of a random cost, for example, ${\rm Var\,}[f(x(\xi),\xi)]\le t$ for some $t$, which is of course  a linkage constraint. As another example, in a supply chain management model (Zhong et al. 2019), a supplier must choose a policy $x$ to satisfy the demand $\xi_i$ for $N$ periods. Let  $f_i(x(\xi),\xi)$ be the allocation function in period $i$. The service standard requires
$$\E(f_i(x(\xi),\xi))\ge \be_i\E(\xi_i)\quad\fo i=1,...,N.$$
That is, the expected proportion of demand from each  period $ i$ that is fulfilled immediately is
at least $\be_i.$ 

\n{\b Decision-dependent distributions.} The probability distribution $\{p(\xi)\}$ in a risk/regret minimization problem may be dependent on the decision made. For example, the probability of demand $\xi$ could be influenced by the advertising decision $x$. Then
$$\min \E(f(x(\xi),\xi))=\min \sum_{\xi\in\Xi}p(x(\xi),\xi)f(x(\xi),\xi)\ \LR\ \min\sum_{\xi\in\Xi}{1\ov|\Xi|}\ba f(x(\xi),\xi),
$$ where $\ba f(x(\xi),\xi)= p(x(\xi),\xi)f(x(\xi),\xi).$ A simple reformulation  will produce   a linkage constraint (possibly, nonconvex).

It can be seen that linkage constraints arise in a wide spectrum of situations. If they are convex, they can be handled by the PHA with a suitable modification, as described in the next section. The PHA may have to be further modified if they are not convex. However,  this is a subject too big to   study in the current paper. Hence, here and below, it is assumed that $h(\cd,\xi)$ is a convex function.

\subsection{A modified PHA that can handle linkage constraints}\label{3.6}

Consider a slightly more general case than a single linkage constraint \reff{cross}, where  our problem is
\beq\label{svi-cross}
\min_{z(\cd)}\cG(z(\cd)) \;\;\h{s.t.}\;\;z(\cd)\in\ba\cC\cap\ba\cN\cap\cS,
\eeq
where $\cS=\{z(\cd):\E(h(z(\xi),\xi))\le t\}$ with $t\in \R^k$ and $h: \R^n\rightarrow\R^k$ with each of its components being a convex function. Obviously, $\E(h(z(\xi),\xi))\le t$ can be written as $\E(\bar{h}(z(\xi),\xi))\le0 $, where $\bar{h}(z(\xi),\xi)=h(z(\xi),\xi)-t. $ By appropriate re-definition of the variables and the functions, without loss of generality, we can simply assume that the linkage constraint is of the form $\E({h}(z(\xi),\xi))\le0 $. Note that by introducing an auxiliary vector $u(\xi)$, the following equivalence holds:
$$\E({h}(z(\xi),\xi))=\sum_{\xi} p(\xi){h}(z(\xi),\xi)\le0 ~\Longleftrightarrow$$
$$\exists u(\xi):~{h}(z(\xi),\xi)\le u(\xi)~\forall~\xi~~\textrm{and}~\sum_{\xi} p(\xi)u(\xi)=0.$$
Thus, enlarging the dimension by setting $\eta(\cdot)=(z(\cdot),u(\cdot))$,  problem \reff{svi-cross} is equivalent to
\begin{eqnarray}\label{enlarge}
&\min\limits_{\eta(\cdot)}& \cG(z(\cdot))\nonumber\\
&\textrm{s.t.}& \eta(\cdot)\in \cC':=\{\eta(\cdot)~|~z(\cdot)\in \bar{\cC},~{h}(z(\xi),\xi)\leq u(\xi)~\forall~\xi\},\nonumber\\
&& \eta(\cdot)\in \NN':=\{\eta(\cdot)~|~z(\cdot)\in \bar{\NN},~\E (u(\xi))=0 \},
\end{eqnarray}
where $\cC'$ is the convex constraint with the decomposable structure with respect to $\xi$, and $\NN'$ is an ``enlarged nonanticipativity subspace" with its complementary subspace being
$$\MM':=\{\lambda(\cdot)=(v(\cdot),w(\cdot))~|~v(\cdot)\in \bar{\MM},~w(\xi)\equiv w~\forall~\xi\}.$$

Therefore, viewing $\NN'$ as the enlarged nonanticipativity space and applying the idea of PHA to solve problem \reff{enlarge}, from $\eta^k(\cdot)\in\NN'$ and $\lambda^k(\cdot)\in\MM'$, i.e.,
$$z^k(\cdot)\in \bar{\NN},~\E (u^k(\xi))=0,~v^k(\cdot)\in \bar{\MM},~w^k(\xi)\equiv w~\forall~\xi,$$
Step 1 is to determine $\hat{\et}^k(\cdot)$ via
\begin{eqnarray*}
\hat{\eta}^k(\xi) =\argmin_{\scriptsize{\begin{array}{c}z\in \bar{C}(\xi)\\{h}(z,\xi)\leq u\end{array}}}\!\l\{ g(z,\xi) -
   v^k(\xi)\cdot z+\frac{r}{2}||z-z^k(\xi)||^2\r.\\
   \l. -w^k(\xi)\cdot u+\frac{r}{2}||u-u^k(\xi)||^2\r\},
\end{eqnarray*}
for every $\xi$. In this case, Step 2 of primal and dual updating turns out to be
$$\eta^{k+1}(\cdot)=P_{\NN'}(\hat{\eta}^k(\cdot))~\Longleftrightarrow~\left\{\begin{array}{l} z^{k+1}(\cdot)=P_{\bar{\NN}}(\hat{z}^k(\cdot)), \vspace{1mm}\\ u^{k+1}(\xi)=\hat{u}^k(\xi)-\E(\hat{u}^k(\xi))~\forall~\xi,\end{array}\right.$$(Note the difference of projections on $\bar{\cN}$ and on the $u$-subspace)
and
$$\lambda^{k+1}(\cdot) = \lambda^k(\cdot)-r P_{\MM'}(\hat{\eta}^k(\cdot))~\Longleftrightarrow~\left\{\begin{array}{l} v^{k+1}(\cdot) = v^k(\cdot)-r P_{\bar{\MM}}(\hat{z}^k(\cdot)), \vspace{1mm}\\ w^{k+1}(\xi)=w^{k}(\xi)-r\E(\hat{u}^k(\xi))~\forall~\xi.\end{array}\right.$$ (Note the differences between updating $v$ and updating $w$.)

In summary, our analysis above leads to the following modified PHA for solving problem \reff{svi-cross}.

\begin{algo}\label{A32}{\b The PHA for regret minimization with linkage constraints}

\n{\b Step 1 (Scenario Decomposition).} Given  $z^k(\cd)\in \ba\NN$, $u^k(\cdot)$ satisfying $\E (u^k(\xi))=0$, $v^k(\cdot)\in \bar{\MM}$ and $w^k(\cdot)$ such that $w^k(\xi)\equiv w~\forall~\xi$. Solve the following optimization problem for each $\xi\in\Xi$ to determine $(\hat{z}^k(\cd),\hat{u}^k(\cd))$.

\begin{eqnarray*}
(\hat{z}^k(\xi),\hat{u}^k(\xi)) =\argmin_{\scriptsize{\begin{array}{c}z\in \bar{C}(\xi)\\{h}(z,\xi)\leq u\end{array}}}\!\l\{ g(z,\xi) -
   v^k(\xi)\cdot z+\frac{r}{2}||z-z^k(\xi)||^2\r.\\
   \l. -w^k(\xi)\cdot u+\frac{r}{2}||u-u^k(\xi)||^2\r\},
\end{eqnarray*}

\n{\b Step 2 (Primal and Dual Update).}
$$z^{k+1}(\cdot)=P_{\bar{\NN}}(\hat{z}^k(\cdot)), ~ u^{k+1}(\xi)=\hat{u}^k(\xi)-\E(\hat{u}^k(\xi))~\forall~\xi,$$
$$v^{k+1}(\cdot) = v^k(\cdot)-r P_{\bar{\MM}}(\hat{z}^k(\cdot)),~w^{k+1}(\xi)=w^{k}(\xi)-r\E(\hat{u}^k(\xi))~\forall~\xi.$$
where $P_{\ba\NN}$ and $P_{\ba\MM}$ are the projection operators to the subspaces $\ba\NN$ and $\ba\MM$, respectively, and $r>0$ is a suitably chosen parameter.
\end{algo}

\begin{theorem}\label{t3}
Suppose that  $g(\cd,\xi)$ is proper, closed, and convex for all $\xi$ and the  problem \reff{svi-cross} satisfies constraint qualification, by which we mean either $\cC'$ is polyhedral and $\cC'\cap\cN'\ne\emptyset$ or $\ri\cC'\cap\cN'\ne\emptyset$). Then the sequence $\{(\eta^k(\cd),\lambda^k(\cd))\}$ generated by \refal{A32} converges to a solution $(\eta^*(\cd),\lambda^*(\cd))$ (if exists at all) of problem \reff{svi-cross}.
Moreover, if $g(x(\cd,\xi)$ is convex piecewise quadratic and $\bar\cC$ and $\cS$ are convex polyhedra, then this sequence converges at $q$-linear rate with respect to the norm
$$\|(\eta^k(\cd)-\eta^*(\cd),\lambda^k(\cd)-\lambda^*(\cd))\|_r.$$
\end{theorem}

Note that the structure of the enlarged nonanticipativity space includes a hyperplane of $u,$ therefore inducing a new dual vector $w.$ It also require a  different update rule for $w$. However, in the same spirit of the proof in Rockafellar and Sun (2019)  we can show that Algorithm 2 is a variety of the proximal point algorithm (Rockafellar, 1976) applied to  a partial inverse problem of Spingarn (1983). Hence we can establish the stated convergence results for the different nonanticipativity constraint. For brevity, we omit the proof.

\section{Numerical results}

All numerical codes are written in MATLAB R2015b and run on a laptop with an Intel(R) Core(TM) i7-7500U 2.70 GHz 2.90 GHz CPU and 16 GB of RAM.

\subsection{An airline seat allocation problem }

We tested a CVaR regret minimization model that arises in an airline seat allocation problem, which is a classical example of real-world application of two-stage stochastic optimization problem. It first appeared in a technical report of London Business School (DeMiguel and Mishra, 2006) and was subsequently used as a subproblem in papers on airline revenue management such as Chen and Homem-de-Mello (2010). For simplicity, consider a single flight with a capacity of 5 units of seats (say, 1 unit = 70 seats) and demand for three fare classes (business, premier, and economy)
with associated revenues $r_B$ = 130; $r_P$ = 100; $r_E$ = 50.
Assume demand arrives in two different stages. In the first stage, there is a deterministic demand of maximal 4 units for economy class seats, and no demand for business or premier class seats. In the second stage, demands of
business and premier classes come in random. The manager is required to allocate the number of seats for each class in each stage to maximize the expected total revenue. Apparently, the dimension of the decision variable is fixed as [1,2], which means that the dimensions of the decision variables in the first and second stage are 1 and 2, respectively, and in this group of experiments, to increase the number of scenarios, we generate the demands for business and premier class seats in the second stage as two random numbers from the discretized normal distribution with mean parameter and standard deviation parameter being $\mu_B=0.9, \sigma_B=0.1$ and $\mu_P=2.3, \sigma_P=0.2$, respectively. The corresponding probability for each scenario is $1/K$, where $K$ is the number of scenarios. For each setting, 10 independent problems are generated. Then, PHA is applied to solve them with $\alpha=0.5$.

For comparison purpose, we also solved the risk-neutral case, which was the original formulation of expectation measure in  DeMiguel and Mishra (2006).

Specifically, the expectation minimization results in the following two-stage stochastic optimization problem:
\begin{eqnarray*}
&\min\limits_{z(\cdot)} & \E[c\cd z(\xi)]\\
& \textrm{s.t.} & A z(\xi)\leq 5,~0\leq z(\xi) \leq D(\xi),~\forall~\xi,\\
&& z(\cdot)\in \NN:=\{z(\cdot)~|~x(\xi)\equiv \textrm{Constant}~\fo \xi\},
\end{eqnarray*}
where $z(\xi)=(x(\xi),b(\xi),p(\xi))^T$, $c=(-50,-130,-100)^T$, $A=(1, 1, 1)$ and $D(\xi)=(4,D_B,D_P)^T$ (recall that ``$T$" stands for the transpose). The corresponding CVaR minimization problem is
\begin{eqnarray*}
&\min\limits_{z'(\cdot)} & \mathbb{E}(y(\xi)+(1-\alpha)^{-1} s(\xi))  \\
& \h{s.t.} & M(\xi)z'(\xi)\leq d(\xi), s(\xi)\ge c^Tz(\xi)-y(\xi)\ \fo \xi,\\
&& 0\leq x(\xi)\leq 4,~0\leq b(\xi) \leq D_B,~0\leq p(\xi) \leq D_P,~s(\xi)\geq 0,\ \fo \xi,\\
&& z'(\cdot)\in \bar{\NN}:=\{z'(\cdot)~|~x(\xi), y(\xi)\equiv \textrm{Constant}~\fo \xi\},
\end{eqnarray*}
where $z'(\cdot)=(y(\cdot),x(\cdot),b(\cdot),p(\cdot),s(\cdot))$,
$$M(\xi)=\l(\begin{array}{ccccc} 0 & 1 & 1 & 1 & 0\\ -1 & -50 & -130 & -100 & -1  \end{array}\r),d(\xi)=\l(\begin{array}{c} 5\\ 0 \end{array}\r).$$

Figure  1 and Table 1 show the performance of PHA for expectation minimization and CVaR minimization with $\alpha=0.5$, when the number of scenarios increases. In Table 1, ``sn" represents the number of scenarios, ``iter" means the average iteration number to convergence for the 10 test problems, ``time(s)" means the average convergence time in seconds for the 10  problems, and ``fval" means their average optimal value.

\begin{figure}[ht]
\centering
\subfigure{\includegraphics[width=0.48\columnwidth]{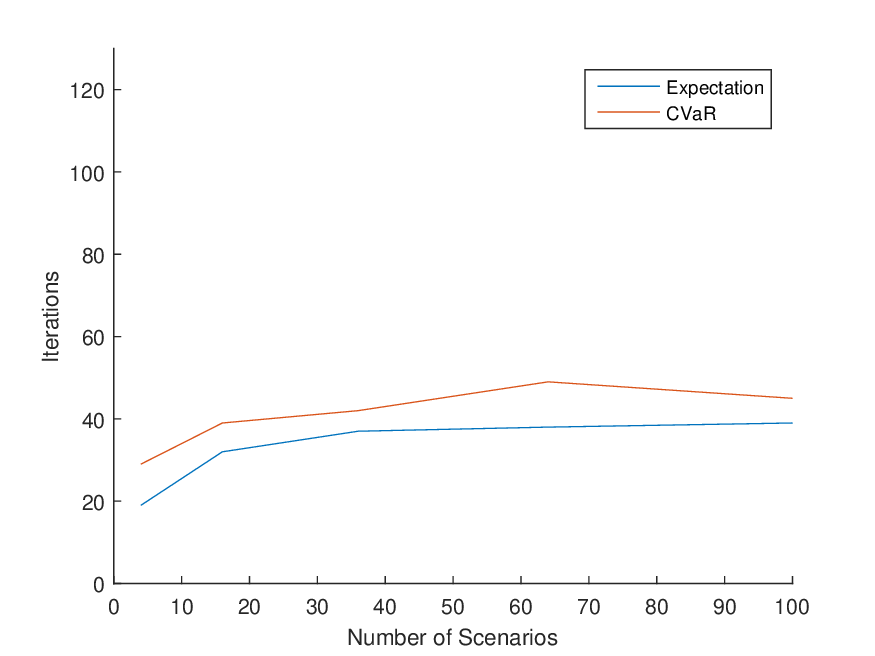}}
\subfigure{\includegraphics[width=0.48\columnwidth]{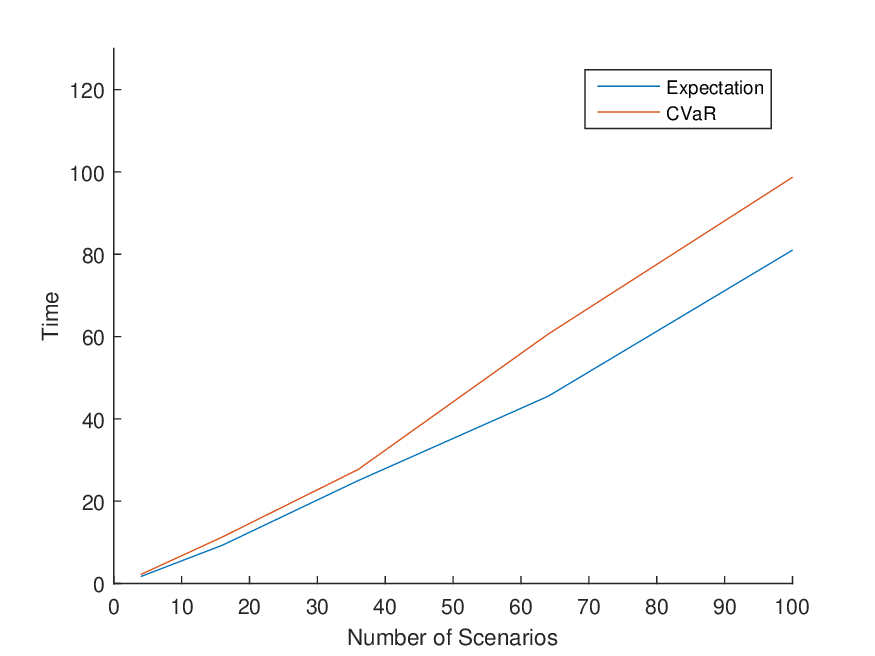}}
\vspace{3mm}\\
\centering{\textbf{Figure  1} Convergence results when sn increases}
\end{figure}

\begin{table}[ht]
\centering

\begin{tabular}{ |c|c|c|c|c|c|c| }
\multicolumn{7}{c}{\textbf{Table 1} Numerical results while sn increases} \vspace{1mm}\\
\hline
\multirow{2}{*}{sn} & \multicolumn{3}{c|}{Expectation} &\multicolumn{3}{c|}{CVaR ($\alpha=0.5$)} \\
\cline{2-7}
& iter & time(s) & fval & iter & time(s) & fval \\
\hline
4	& 19 &	1.7 &	-433.347 &	29 & 2.2 &	-426.344 \\
16	& 32 &	9.3 &	-431.661 &	39 & 11.3 &	-424.622 \\
36  & 37 &	25.0 &	-429.710 &	42 & 27.7 &	-422.977 \\
64	& 38 &	45.5 &	-430.192 &	49 & 60.6 &	-423.500 \\
100 & 39 &	81.0 &	-430.389 &	45 & 98.7 &	-423.664 \\
\hline
\end{tabular}
\end{table}

It can be seen from Figure  1 and Table 1 that the number of iterations grows slowly and time to convergence grows at linear rate for CVaR minimization  when the number of scenarios increases. It takes more iterations and hence more time to convergence for CVaR minimization than  expectation minimization. Since CVaR is risk-averse, the cost of it is higher.

\subsection{Numerical results for general two-stage linear risk/regret minimization}

In order to test the efficiency of PHA for regret minimization, we  test a series of randomly generated two-stage linear risk minimization problems
\begin{eqnarray}\label{2stage-linear}
\min_{x(\cd)\geq 0}\;\cR(f(x(\xi),\xi)),
\end{eqnarray}
where $f(x(\xi),\xi)$ is the optimal value function
$$f(x(\xi),\xi):=\min_{x_2(\cd)\geq 0}\big\{ q^T x_1(\xi)+c(\xi)^T x_2(\xi)~:~A(\xi)x_1(\xi)+B(\xi)x_2(\xi)=d(\xi)\big\}.$$
The corresponding regret minimization problem is
\begin{eqnarray}\label{2stage-linear-regret}
&\min\limits_{x(\cdot),y\in \R} & y+\VV\big(q^Tx_1(\xi)+c(\xi)^Tx_2(\xi)-y\big)\nonumber\\
&\h{s.t. }& A(\xi)x_1(\xi)+B(\xi)x_2(\xi)=d(\xi),~\fo \xi\in\Xi,\\
&& x_1(\xi)\geq 0,~x_2(\xi)\geq 0,~\fo \xi\in\Xi,\nonumber\\
&& x(\cdot)\in \NN:=\{x(\cdot)=(x_1(\cdot),x_2(\cdot))~|~x_1(\xi)\equiv\textrm{Constant}~\fo \xi\}.\nonumber
\end{eqnarray}
 As mentioned in Section 3.1,  it is often more convenient to handle regret minimization than to handle risk minimization directly. Therefore, our numerical test is devoted to problem \reff{2stage-linear-regret} for CVaR.

The number of scenarios is fixed at 20 and the dimension of the decision variable $x$ rises from [10,10] to [50,50] in the two stages, with randomly generated $A(\xi), B(\xi),c(\xi)$ and $d(\xi)$.
For each setting, we  generate 10 random problems and use PHA to solve the expectation minimization problems and the CVaR minimization problems with $\alpha=0.5$, respectively.

Figure  2 and Table 2 show the performance of PHA for expectation minimization and that of the modified PHA for CVaR minimization. From Figure 2, it is easy to see that the number of iterations to convergence grows steadily when the dimension of problems increases, while convergent time grows at faster rate for both expectation minimization and CVaR minimization ($\alpha=0.5$). In addition, it seems that it takes much more iterations and time  for CVaR minimization to converge  than the expectation minimization, especially when problem dimension gets large.

\begin{figure}[ht]
\centering
\subfigure{\includegraphics[width=0.48\columnwidth]{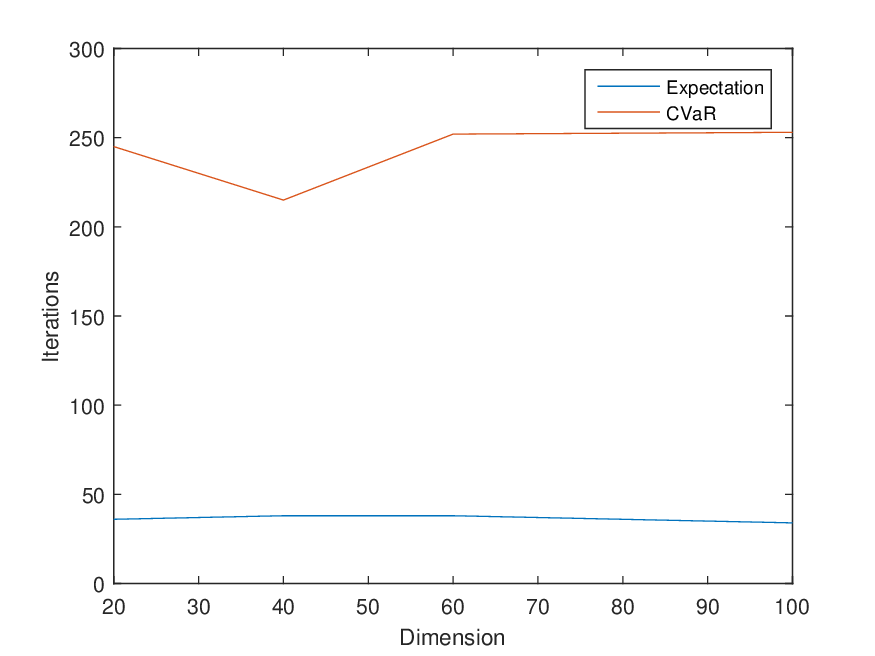}}
\subfigure{\includegraphics[width=0.48\columnwidth]{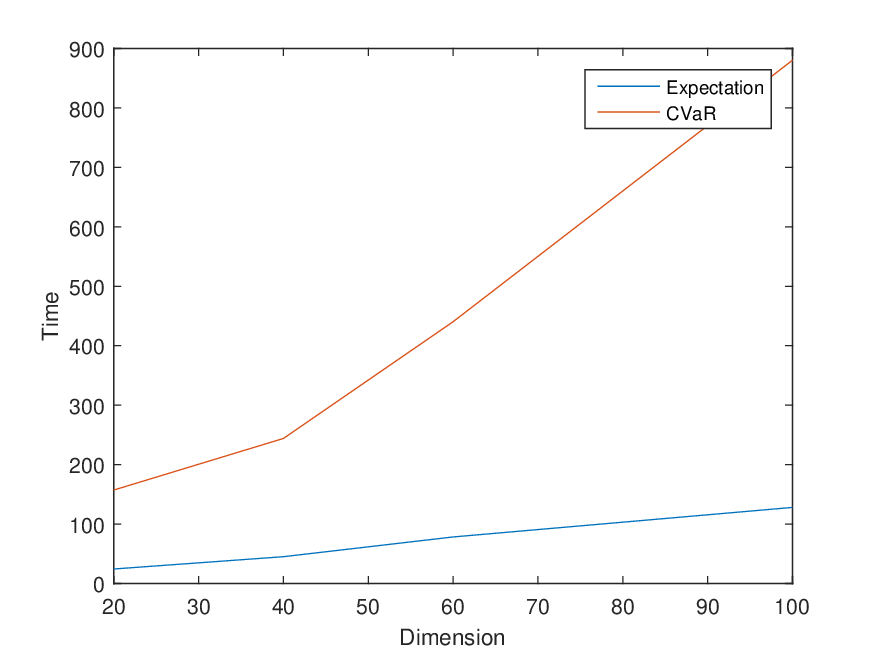}}
\vspace{3mm}\\
\centering{\textbf{Figure  2} Convergence results when dimension increases}
\end{figure}

\begin{table}[ht]
\centering
 \begin{tabular}{ |c|c|c|c|c|c|c| }
\multicolumn{7}{c}{\textbf{Table 2} Numerical results while dim increases (sn=20)} \vspace{1mm}\\
\hline
\multirow{2}{*}{dim} & \multicolumn{3}{c|}{Expectation} &\multicolumn{3}{c|}{CVaR ($\alpha=0.5$)} \\
\cline{2-7}
& iter & time(s) & fval & iter & time(s) & fval \\
\hline
$[10,10]$ & 36 & 24.4   & 2.52  & 245 &	157.2 &	3.38 \\
$[20,20]$ & 38 & 45.1	& 4.99	& 215 & 244.0 & 6.25 \\
$[30,30]$ & 38 & 78.3	& 7.49	& 252 & 440.4 & 8.86 \\
$[50,50]$ & 34 & 128.0  & 12.49	& 253 & 880.3 & 14.26\\
\hline
\end{tabular}
\end{table}

\section{Conclusion}

A dual relationship between risk measure and regret measure is established. It helps to build a list of correspondences  between useful coherent and averse risk and regret measures. Based on such dual representation of risk measure, the multistage risk minimization problem can be converted to a multistage regret minimization problem. A progressive hedging algorithm is proposed for solving the corresponding  minimization problems. In case that linkage constraints arise in the risk and regret minimization problem, the progressive hedging algorithm can be modified to take advantage of the hidden decomposability  of the problems. Preliminary numerical results are reported to show the efficiency of the progressive hedging algorithms for risk-neutral and risk-averse practical or randomly generated problems.

\begin{acknowledgements}
The research of Sun is partially supported by Australian Research Council (Grant DP160102819). The research of Yang is partially supported by National Natural Science Foundation of China (Key Grant 11431004). The research of Yao is partially supported by National Natural Science Foundation of China (Grants 11671145 and B14019). The research of Zhang is partially supported by Chinese Academy of Science (Grant Y932231).
\end{acknowledgements}

\end{document}